# They Came From the Deep in the Supernova:
## The Origin of TiC and Metal Subgrains in Presolar Graphite Grains

Katharina Lodders [1]

**Abstract**

A new formation scenario for TiC and Fe,Ni-metal inclusions in presolar graphite grains of supernova origin is described. The mineralogy and chemistry require condensation of (Fe,Ni)-titanides from Fe, Ni, and Ti-rich gaseous ejecta, subsequent carburization to make TiC and metal, and encapsulation into graphite. Titanides only condense if Si is depleted relative to heavier mass elements, which requires $\alpha$-rich freeze-out and a deep mass-cut for the supernova ejecta. This Si-poor core material must remain unmixed with other supernova zones until the titanides condensed. This can be accomplished by transport of core ejecta in bipolar jets through the major expanding supernova zone ejecta. If the jets stall in regions dominated by C-rich ejecta such as C-He zone where graphite condenses, thermochemically favored *in situ* carburization of the titanides - either before or during encapsulation into condensing graphite - leads to a TiC and metal composite. This scenario agrees with theoretical models and observations of asymmetric core collapse in supernovae that are associated with bipolar jets loaded with iron-peak elements.

*Subject headings:* astrochemistry --- nucleosynthesis --- supernovae: general --- supernova remnants

[1] Planetary Chemistry Laboratory, Dept. of Earth & Planetary Sciences, and McDonnell Center for the Space Sciences, Washington University, Campus Box 1169, St. Louis, MO 63130, USA; lodders@wustl.edu

## 1. Introduction

A small fraction of the known presolar graphite grains is thought to originate in the ejecta of core-collapse, type II supernovae, SNII, (e.g., reviews by Zinner, 1998, Lodders & Amari 2005). These micron-size presolar graphite grains may contain 10-500 nm-size TiC and 10-80 nm-size kamacite (Fe-rich) metal subgrains (Croat et al. 2003, henceforth C03). The FeNi-metal subgrains are almost exclusively attached to TiC subgrains (only 3 out of 84 metal grains were not - see C03). Crystallographic evidence for epitaxial growth of FeNi-metal onto many TiC subgrains suggests the condensation sequence TiC-metal-graphite (C03).

However, it is not easy to only condense TiC, metal, and graphite in this order without any Si compounds. Usually, Si is as abundant as Fe (except in the innermost SN zones) so abundant Si-compounds condense. However, SNII graphites *typically* do not contain internal SiC grains or other Si-rich phases. The graphite has only ~1 wt% Si (Hoppe et al. 1995, C03), TiC subgrains only have ~5 at% Si on average, and Si is usually not detected in metal subgrains (C03). The low Si abundances rule out condensation of Si-bearing phases before TiC, metal, and graphite.

In contrast to the SNII graphites, presolar graphite grains from AGB stars may contain TiC and refractory carbides of s-process elements, and typically, they have no metal and SiC subgrains (Bernatowicz et al. 1991, 1996). This agrees with the TiC-graphite-SiC-silicide condensation sequence in C-rich circumstellar shells of AGB stars with otherwise solar-like major element composition (e.g., Lodders & Fegley 1995, 1999, Bernatowicz et al. 1996), but condensation from such a gas does not explain the SNII graphites with internal TiC and metal subgrains. However, elemental abundances in SNII ejecta are altered from solar by nucleosynthesis. The condensation sequences in different SNII zones are calculated in §§ 2 & 3. Condensation from any individual SNII zone cannot account for the observations. A condensation-reaction model for the TiC-metal-graphite grains is proposed in § 4. Section §5 briefly describes that asymmetric core collapse SNII models and astronomical observations are consistent with the proposed condensation model summarized in §6.

## 2. Condensates from Supernova Zone Compositions

Condensation calculations were done for averaged compositions of individual SNII zones from the $25M_\odot$ supernova model S25P by Rauscher et al. (2002, henceforth R02[2]). This more energetic model has $0.2M_\odot$ $^{56}$Ni in the ejecta (twice that of their conventional $25M_\odot$ SNII model). This is more favorable for the required Ni and Fe-rich chemistry of the presolar graphites with internal TiC and metal grains. Condensation results for conventional $15M_\odot$ and $25M_\odot$ SNII models by R02 are similar but only results for model S25P are given here.

Calculations were done with the CONDOR code (Lodders 2003). A nominal total pressure of $P_{tot} = 10^{-7}$ bar was adopted from the estimates by C03, but calculations were also done for different $P_{tot}$. The condensates for the different SNII zones are broadly similar to those reported by Lattimer et al. (1978) and Ebel & Grossman (2000) except that here, silicides and titanides are included. Thermochemical properties for these are from Barin (1989), Knacke et al. (1991), and Kubaschewski (1983).

Compositions of the nominal SNII zones, named by their 2 most abundant elements, are in Table 1. The He-H, He-N, O-C, O-Ne and O-Si zones with C/O<1 give condensates such as corundum, hibonite, perovskite, Mg-silicates, and metal. Therefore, these zones are not considered further as contributors to the SNII graphites with subgrains.

The inner Si-S and Fe-Ni zones are very C- and O-poor (Table 1) and condensates are silicides (Ti$_5$Si$_3$, (Fe,Ni)Si, (Fe,Ni)$_3$Si,), sulfides, and metal (Table 2). Iron and Ni always co-condense so it does not matter whether $^{56}$Ni has decayed into $^{56}$Fe or not. Graphite and TiC are not among the refractory condensates. The SNII graphites studied by C03 are Si-poor and without Si-rich subgrains, so the chemistry from the Si-S and Ni-Fe-zones seems irrelevant.

---

[2] See also: http://www.ucolick.org/~alex/data/

Only the C-He zone with high C/O favors graphite formation but condensation solely from this zone cannot explain the observation of graphite with internal TiC and metal. Graphite condenses several 100 K higher than TiC at $10^{-7}$ bar (Table 2). Changes in $P_{tot}$ by a few orders of magnitude do not alter this conclusion much, and it is difficult to reverse graphite and TiC condensation. If mixing of C-He zone material with material from its neighboring, O-rich zones occurs, the high C/O of ~13 that causes the high graphite condensation temperature, can be reduced. Mixing of material from different zones is required to explain the C, N, and O isotopic compositions in presolar SNII graphites (e.g., Travaglio et al. 1999). As long as mixing gives C/O >1, TiC and graphite can condense and their condensation temperatures reverse at certain $P_{tot}$. But even if particular C/O ratios and total pressures lead to TiC condensation prior to graphite, there is condensation of $(Fe,Ni)_3Si$ (unobserved) instead of FeNi metal (observed). This occurs whether mixing with neighboring zones reduces the original C/O. Also, SiC condenses prior to silicide in the C-He zone. The observed TiC-metal-graphite condensation sequence is incompatible with the calculated sequence graphite-TiC-SiC-$(Fe,Ni)_3Si$ for the C-He zone, or TiC-graphite-SiC-$(Fe,Ni)_3Si$ for the C-He zone with admixtures from the O-Ne and/or He-N zone.

None of the zone-averaged SNII compositions gives the condensation sequence TiC-metal-graphite implied by the observations. However, presolar SNII graphite grains also contain decay products of $^{44}$Ti ($t_{1/2}$ =58 yr) and $^{49}$V ($t_{1/2}$ =337 d), and their production is expected only in the innermost SNII zones (Nittler et al. 1996, Hoppe & Besmehn 2002). As the grains do not contain much Si, Si-poor compositions are needed. All this leads us to dig deeper into the Ni-core.

## 3. Condensates from Si- and S-poor Ni-Fe zone compositions

I now consider the effect of compositional gradients across the Ni-Fe-zone. Figure 1a shows the mass fractions of Fe, Ni, Ti, Si, Ca and S as a function of interior SNII mass from the S25P model by R02. The Ni-Fe-zone composition depends on the energy of the reverse shock, mass-cut, and fall back from the SNII explosion. The border between the Si-S and the Ni-Fe-zones is taken where mass fractions of Fe (after $^{56}$Ni decay) are larger than that of Si. From this border inwards, Fe and Ni abundances increase to constant high values, whereas Si, S, and Ca abundances significantly drop, go through a minimum and level out to constant low values. The Ti abundance increases less steeply than that of Fe and Ni, and moves through a minimum like Si, Ca, and S before leveling out. The steep changes over orders of magnitude show that Ca, Si, and S abundances of the averaged Ni-Fe-zone in Table 1 are mainly given by Ni-Fe-zone layers near the Si-S zone. Only deep in the Ni-Fe-zone are Fe, Ni, and Ti much more abundant than Si.

Figure 1b shows the condensation chemistry along the mass profile in Figure 1a. Compositions near the Si-S zone give the same condensates ($Ti_5Si_3$, $(Fe,Ni)Si$, $(Fe,Ni)_3Si$, Fe,Ni-metal, and CaS) as in the average Ni-Fe-zone (Table 2). But deeper, the titanide $(Fe,Ni)_2Ti$ condenses instead of the silicides because Si abundances are minimized. After the Si minimum is passed, initial condensates are $Ti_5Si_3$, $(Fe,Ni)_2Ti$, and FeNi. Condensation of $(Fe,Ni)_2Ti$ is the key to the presolar grain observations (§4), but it is not the first condensate for the compositions in Figure 1a and CaS or $Ti_5Si_3$ – neither observed in the presolar graphites – condenses first (Figure 1b). Although the Ni-Fe-zone in the S25P model by R02 is already more Fe and Ni rich than in their conventional models, there still is abundant S, Si, and Ca from incomplete Si-burning. This changes under very energetic conditions when complete Si-burning with α-rich freeze-out leaves heavy Fe-peak nuclides (including $^{44}$Ti) and significantly reduces Si, S, and other lighter Z elements, which is required to explain the observed chemistry.

Figure 2 shows the changes when Si and S abundances in the nominal Ni-Fe-zone composition (Table 1) are decreased by the same factor. The nominal composition gives $Ti_5Si_3$, $(Fe,Ni)Si$ which transforms into $(Fe,Ni)_3Si$ at lower temperatures, and FeNi metal. With decreasing Si and S content, metal condensation temperatures are unchanged but those of the silicides drop dramatically. The Fe, Ni-bearing silicides are replaced by (Fe,Ni)$_2$Ti when Si and S abundances are ~4% of their nominal values. The (Fe,Ni)$_2$Ti condensation temperature increases with further Si and S reductions. Decreasing the Si and S abundances by more than a factor of 1000 only leaves (Fe,Ni)$_2$Ti and FeNi metal.

## 4. Formation of the TiC-metal composites

Condensation of (Fe,Ni)$_2$Ti (and/or (Fe,Ni)Ti, depending on details of $P_{tot}$ and compositions) from Ni-Fe-zone ejecta opens a new route to make the TiC and metal subgrains in presolar graphites. Consider that an Fe,Ni-titanide condensed from Ni-Fe-zone ejecta and reached the C-rich C-He zone. There thermochemically favored *in situ* carburization of the titanide gives a TiC-metal composite:

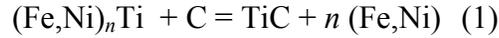

$$(Fe,Ni)_n Ti + C = TiC + n (Fe,Ni) \quad (1)$$

This reaction may happen before or during encapsulation of the particles into graphite as it condenses and fits the observations by C03 that almost every metal subgrain is attached to a TiC subgrain in the graphites. Depending on temperature, titanides and metal are either solid or liquid (melting points 1427°C (Fe$_2$Ti), 1570°C (FeTi), 1538°C (Fe)) whereas TiC (mp. 3067°C) is solid. Liquids can facilitate diffusion of C to form TiC from the titanides. The observed crystallographic alignments of TiC and metal surfaces in the presolar graphite grains may result from solid (or liquid) phase reordering during *in situ* carburization, instead of epitaxial growth (i.e., metal-vapor deposition onto TiC) suggested by C03. However, this requires more investigation.

*In situ* titanide carburization is consistent with the observed diameters of the TiC-metal assemblages in the presolar graphites. The TiC grain diameters typically are larger than those of their attached metal grains and TiC/metal diameter ratios in the sliced graphite grains range from ~2 to ~6 (C03). If metal were freely condensing onto TiC, one should expect much thicker metal layers than TiC diameters because Fe and Ni abundances, and thus the condensable metal volume, are generally much larger than that of Ti in any SNII zone.

It is easy to compare expected TiC/metal diameter ratios with the presolar grain observations. In reaction 1, TiC and FeNi-metal form in molar proportions of 1:2 from (Fe,Ni)$_2$Ti ($n$=2) or 1:1 from (Fe,Ni)Ti ($n$=1). Using densities of $\rho_{Fe}$ = 7.87 and $\rho_{TiC}$ = 4.93 cm$^3$/g, the molar volume ratio is $V_{TiC}/V_{met}$ = 0.86 for $n$=2 (1.7 for $n$=1). The Ni content in the metal is neglected for simplicity. Using denser FeNi alloys slightly increases the TiC/metal diameter ratios. If the assemblage consists of a TiC sphere next to a metal sphere, the diameter ratio is $a_{TiC}/a_{met}$ = 0.95 for $n$=2 (1.2 for $n$=1). The metal thickness ($x_{met}$) on *one* side of a solid TiC sphere evenly coated by metal (as seen when the sphere is sliced) relative to the TiC diameter is $a_{TiC}/x_{met}$ = 6.8 for $n$=2 (12 for $n$=1). The calculated TiC/metal diameter ratios of ~1 to ~7 from (Fe,Ni)$_2$Ti carburization are similar to the observed ratios of ~2 to ~6. This suggests condensation and subsequent carburization of (Fe,Ni)$_2$Ti instead of TiC and metal condensation from a gas enriched in Fe and Ni.

## 5. Supernova Setting

The proposed formation process of internal TiC and metal grains in presolar SNII graphite requires: (1) Complete α-freeze-out near the SNII core to make Fe, Ni and Ti-rich but Si and S-poor ejecta. (2) Transport of Ni-Fe-zone ejecta across the Si-S, O-Si, O-Ne, and O-C zones without mixing so that Fe,Ni-titanides condense. (3) Carburization of the titanide in the C-He zone. These conditions can be realized by jet-driven, non-spherical core-collapse supernovae.

Khokhlov et al. (1999) showed that magneto-rotational induced supersonic jets in a core of a massive star cause non-spherical SN explosions in which most of the material is expelled in oblate, distorted ejecta, and a smaller amount is ejected in high-velocity bipolar jets. Nagataki et al. (1997) and Nagataki (2000) showed that jet-like explosions cause active α-rich freeze-out behind the more energetic shock wave in the polar regions, and that the high-velocity jets launched from the poles are very likely rich in Fe-peak elements. Hence, bipolar jets from asymmetric core collapse provide the nucleosynthesis setting for α-rich freeze-out and the transport of Fe, Ni, and Ti-rich ejecta to the C-He zone to make the observed TiC-metal composites in the presolar graphites.

Non-spherical explosion models with associated jets can account for many observations in SN and their remnants (e.g., SN 1987A, Cas A, Vela) where conventional spherical models fail (e.g., Nagataki et al. 1997, Nagataki, 2000, Khokhlov et al. 1999, Hoeflich et al. 2003, Hungerford et al. 2003). For example, double-peaked SN light curves fired by $^{56}$Ni-containing ejecta moving at different velocities, and asymmetric features in supernova remnants are most plausibly due to high-velocity bipolar jets that cause $^{56}$Ni- (later Fe)-rich "bullets", "clumps", and "knots" in the outer ejecta zones (see, e.g., Wang et al. 2002, Hwang & Laming 2003, Willingale et al. 2003, Tominaga et al. 2005). Wang et al. (2002) concluded that the observed bipolar structure of SN1987A and the high-velocity ejecta rich in $^{56}$Ni are best explained by assuming that non-relativistic jets drove the SN explosion and led to $^{56}$Ni and $^{44}$Ti-rich plumes ejected along the jet axis to the overall expanding ejecta. It seems that many SN observations almost directly probe the necessary conditions to make the TiC-metal composites in the presolar graphites.

## 6. Summary

Presolar SNII graphite with composite TiC-metal subgrains cannot be formed from any single SNII zone composition such as the C-rich C-He zone, nor by considering moderate mixing of the C-He zone with its neighboring He-N and/or O-C zones. The formation of TiC-metal composites requires complete Si-burning with α-rich freeze-out to obtain ejecta from which $(Fe,Ni)_2Ti$ can condense. Jet-driven, axisymmetric core collapse can lead to α-rich freeze-out in the polar regions at the Ni-core, and cause bipolar jets carrying ejecta rich in Fe-peak elements (e.g., Nagataki et al. 1997, Khokhlov et al. 1999). Condensation of $(Fe,Ni)_2Ti$ must occur in the chemical environment of the jets that pass through the main ejecta. If the jets stall in the C-rich C-He zone, carburization of the titanide to the TiC-metal composite takes place either before or during encapsulation of the TiC-metal composite into graphite.

I thank T. Bernatowicz and K. Croat for presolar grains discussions and B. Fegley and S. Woosley for comments on the manuscript. Careful referee comments by R. Gallino are much appreciated. Work supported in part by NASA grant NNG04GG13G.

Table 1. Adopted Atomic Zone Abundances of a 25M$_\odot$ SN [a]

| Zone | Fe-Ni | Si-S | O-Si | O-Ne | O-C | C-He | N-He | He-H |
|---|---|---|---|---|---|---|---|---|
| H | 235 | 0.421 | 0.056 | 21 | 25 | 0.564 | 5.06e9 | 2.03e10 |
| He | 1.09e7 | 90 | 40 | 568 | 3.79e8 | 9.28e9 | 8.44e9 | 4.63e9 |
| C | 35 | 111 | 2.20e4 | 2.80e6 | 6.94e8 | 4.38e7 | 1.02e6 | 3.09e6 |
| N | 9.8 | 6.78 | 137 | 4190 | 3.20e4 | 3.10e5 | 3.38e7 | 2.10e7 |
| O | 190 | 2.68e5 | 6.81e6 | 1.04e8 | 1.42e9 | 3.21e6 | 2.18e6 | 1.29e7 |
| Mg | 820 | 588 | 1.99e5 | 6.21e6 | 2.77e7 | 1.23e6 | 1.06e6 | 1.07e6 |
| Al | 540 | 575 | 2.42e4 | 6.43e5 | 1.98e5 | 1.07e5 | 1.03e5 | 8.68e4 |
| Si | 1.00e6 | 1.00e6 | 1.00e6 | 1.00e6 | 1.00e6 | 1.00e6 | 1.00e6 | 1.00e6 |
| S | 7.96e5 | 4.62e5 | 2.70e5 | 3.05e4 | 2.48e5 | 4.84e5 | 5.15e5 | 5.15e5 |
| Ca | 2.60e5 | 4.49e4 | 9.19e3 | 1.12e3 | 1.88e4 | 5.46e4 | 6.11e4 | 6.11e4 |
| Ti | 1.49e4 | 367 | 260 | 410 | 4660 | 2360 | 2400 | 2400 |
| Fe | 9.62e6 | 7.69e4 | 1150 | 1.49e4 | 2.70e5 | 8.60e5 | 9.00e5 | 9.00e5 |
| Ni | 5.90e5 | 2520 | 2020 | 2.53e4 | 3.68e5 | 4.81e4 | 4.94e4 | 4.94e4 |
| C/O | 0.18 | 4.2e-4 | 3.2e-3 | 0.027 | 0.49 | 13.6 | 0.47 | 0.24 |

[a] scaled to Si=10$^6$ atoms, only major elements listed here. Data from model S25P by R02

Table 2. Condensates and Condensation Temperatures[a]

| Zone: | Fe-Ni | Si-S | C-He |
|---|---|---|---|
| C | - | - | 2055 |
| TiC | - | - | 1579 |
| SiC | - | - | 1405 |
| Ti$_5$Si$_3$ | 1622 | 1641 | - |
| TiSi | - | 1609 | - |
| (Fe,Ni)Si | 1528 | 1546 | - |
| (Fe,Ni)$_3$Si | 1461 | - | 1176 |
| Fe,Ni | 1434 | - | - |
| Si | - | 1507 | - |
| CaS | 1398 | 1262 | 1099 |

[a] at 10$^{-7}$ bar and averaged zone compositions in Table 1.

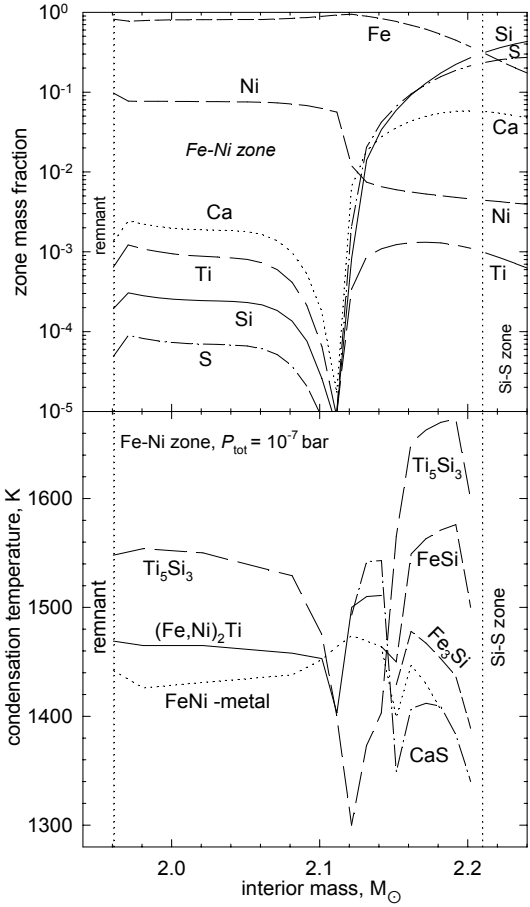

Figure 1a. (top) The Ni-Fe-zone composition along the interior mass for the $25_\odot$ SNII model S25P by R02. Vertical lines indicate borders of the stellar remnant and the neighboring Si-S zone. 1b. (bottom) Condensates and condensation temperatures at $P_{tot} = 10^{-7}$ bar for the compositions in Fig. 1a. At M <2.1 $M_\odot$ CaS condensation temperatures are below 1300 K because Ca and S abundances drop (Fig. 1a).

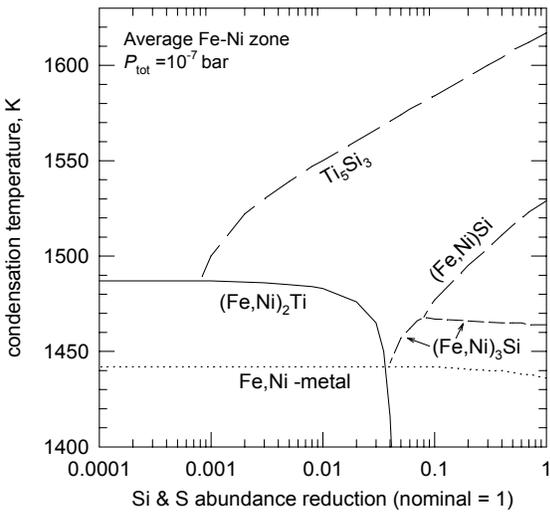

Figure 2. The effect on condensates of reducing Si and S abundances by the same factor from the nominal Ni-Fe-zone composition (Table 1). The CaS condensation temperatures (not shown) decrease from ~1400 K at the nominal composition with decreasing S abundances.